\documentclass[aps,twocolumn,pra,
preprintnumbers,amsmath,amssymb,superscriptaddress]{revtex4}
\usepackage{graphicx}
\usepackage{color}
\usepackage{epstopdf}

\definecolor{refkey}{rgb}{0.9451,0.2706,0.4941}
\definecolor{labelkey}{rgb}{0.9451,0.2706,0.4941}
 
\newcommand{\del} {\partial}
\newcommand{\eq} {equation}
\newcommand{\eqa} {eqnarray}
\newcommand{\NN} {\nonumber}
\newcommand{\sgn}{{\rm sgn}}

\begin{document}

\preprint{WIS/05/16-APR-DPPA}

\title{How to resum perturbative series
in 3d $\mathcal{N}=2$ Chern-Simons matter theories}

\date{\today}

\author{Masazumi Honda}\email[]{masazumi.honda@weizmann.ac.il} 
\affiliation{{\it Department of Particle Physics, Weizmann Institute of Science, Rehovot 7610001, Israel}}

\begin{abstract}
Continuing the work arXiv:1603.06207,
we study perturbative series 
in general 3d $\mathcal{N}=2$ supersymmetric Chern-Simons matter theory
with $U(1)_R$ symmetry,
which is given by a power series expansion of inverse Chern-Simons levels.
We find that
the perturbative series are usually non-Borel summable along positive real axis for various observables.
Alternatively
we prove that
the perturbative series are always Borel summable 
along negative (positive) imaginary axis 
for positive (negative) Chern-Simons levels.
It turns out that
the Borel resummations along this direction are the same as exact results 
and
therefore correct ways of resumming the perturbative series.
\end{abstract}

\maketitle



\noindent

\section{Introduction}
Chern-Simons (CS) theories coupled to matters
play important roles in high energy physics and condensed matter physics.
When CS levels are finite,
the theories are strongly coupled and 
systematic analysis is restricted.
While one can always setup perturbation theories in the CS theories
by expanding observables around infinite CS levels,
the perturbative series are usually divergent
as in typical interacting quantum field theory (QFT) \cite{Dyson:1952tj}.
Therefore 
it is generically hard to obtain information on the strongly coupled systems
from the perturbative series.
In this paper
we address this problem and
discuss that
we can obtain exact results
by appropriately resumming the perturbative series 
in 3d $\mathcal{N}=2$ supersymmetric (SUSY) CS matter theories.

One of standard methods to resum divergent series
is Borel resummation.
Given a perturbative series 
$\sum_{\ell =0}^\infty c_\ell g^{a+\ell} $
of a quantity $I(g)$,
its Borel resummation along the direction $\theta$ is defined by
\begin{\eq}
\mathcal{S}_\theta I (g)
=\int_0^{\infty e^{i\theta}} dt\ 
e^{-\frac{t}{g}} \mathcal{B}I(t) .
\label{eq:Borel} 
\end{\eq}
Here $\mathcal{B}I(t)$ is analytic continuation of 
the formal Borel transformation  
$\sum_{\ell =0}^\infty 
\frac{c_\ell}{\Gamma (a+\ell )} t^{a+\ell -1}$
after performing the summation.
While
perturbative series in typical interacting QFT is 
expected to be non-Borel summable along positive real axis
due to singularities 
in $\mathcal{B}I(t)$ \cite{'tHooft:1977am},
it is natural to ask
when perturbative series is Borel summable 
along $\mathbb{R}_+$
and if it is non-Borel summable, 
what is a correct way to resum the perturbative series.
This is not just a technical question but physically fundamental question
since this is related to how to define interacting QFT's.

In \cite{Honda:2016mvg}
the author initiated to address this question.
We have proven that
perturbative series 
in 4d $\mathcal{N}=2$ and 5d $\mathcal{N}=1$ SUSY gauge theories 
with Lagrangians
are Borel summable along positive real axis for various observables \footnote{
See \cite{Russo:2012kj,Gerchkovitz:2016gxx} for earlier checks of this result in few examples.    
}.
This result for the 4d $\mathcal{N}=2$ theories
is expected from a recent proposal 
on a semi-classical realization of infrared renormalons \cite{Argyres:2012vv},
where the semiclassical solution does not exist 
in the $\mathcal{N}=2$ theories (see also \cite{Dunne:2012ae}).
Then it is natural to apply the technique in \cite{Honda:2016mvg}
to another class of theories.
In this paper
we study perturbative series 
in general 3d $\mathcal{N}=2$ SUSY CS matter theories
with $U(1)_R$ symmetry in terms of inverse CS levels \footnote{
Note that
we consider 
resummation of $1/k$-expansion with fixed rank 
and this is not one of $1/N$-expansion
studied in the context of M2-brane theories \cite{Grassi:2014cla}.
} (see \cite{Russo:2012kj} for studies of 3d $\mathcal{N}=6$ case).
We apply the technique in \cite{Honda:2016mvg}
to localization formula \cite{Pestun:2007rz} for various observables
in 3d $\mathcal{N}=2$ CS matter theories.

Nevertheless
we find highly different results 
from the 4d $\mathcal{N}=2$ and 5d $\mathcal{N}=1$ theories.
First of all
we find that
perturbative series 
are usually
{\it not} Borel summable 
along $\mathbb{R}_+$
for various observables.
Alternatively
we prove that
the perturbative series are always Borel summable 
along negative imaginary axis for positive CS levels
and positive imaginary axis for negative CS levels.
We also prove that
the Borel resummations along this direction are the same as exact results \footnote{
We assume that
the observables are well-defined
though ill-defined cases are also interesting \cite{Morita:2011cs}.
}.
Our main result is schematically written as
(more precise statement is \eqref{eq:mainO})
\begin{\eq}
\mathcal{O}(g)
=\mathcal{S}_{\frac{-\pi \sgn (k)}{2}}  \mathcal{O}(g)
= \int_0^{-i\sgn (k) \infty} dt\ e^{-\frac{t}{g}} \mathcal{B}\mathcal{O}(t) ,
\label{eq:conclusion}
\end{\eq}
where $g\propto 1/|k|$ with CS level $k$ and 
$\mathcal{B}\mathcal{O}(t)$ is Borel transformation \footnote{
We simply refer to analytic continuation of formal Borel transformation
as Borel transformation below.
}
of small-$g$ expansion of the observable $\mathcal{O}(g)$.
This means that
exact results are given by
the Borel resummations along the direction $\theta =-\pi \sgn (k)/2$.
In sec.~\ref{sec:derivation}
we proove the results \eqref{eq:conclusion} 
for $S^3$ partition function,
SUSY Wilson loops, Bremsstrahrung function, 
two-point function of $U(1)$ flavor symmetry currents,
partition function on on squashed lens space
and two-point function of stress tensor.

Our rerult \eqref{eq:conclusion} is quite surprising in the following reason.
When the perturbative series are not Borel summable along $\mathbb{R}_+$,
we usually consider a possibility of cancellations of 
the perturbative ambiguities by contributions from other saddle points
such as instantons
or perform more complicated analysis such as median resummation
to find a correct integral contour.
We find that
we can skip the complicated analyzes and 
directly find the correct integral contour
though understanding from the usual analyzes should be important.
We expect that
our result is very important
also for understanding non-SUSY CS matter theories.
While we do not know 
if the perturbative series in the non-SUSY theories 
are Borel summable along the contour in \eqref{eq:conclusion},
it is natural to expect that
this choice of the contour makes analysis highly simplified \footnote{
For instance, 
if we consider a non-SUSY theory regarded as
a continuum deformation of the 3d $\mathcal{N}=2$ CS matter theories,
then observables in this theory would approximately satisfy \eqref{eq:conclusion}
for small deformation parameters.
}.

\section{Derivation of results}
\label{sec:derivation}
\subsection{Partition function on $S^3$}
Suppose 3d $\mathcal{N}=2$ CS matter theory 
with a semi-simple gauge group $G=G_1 \times\cdots\times G_n$,
which is coupled to
chiral multiplets of representations $(\mathbf{R_1} ,\cdots ,\mathbf{R_{N_f}} )$ with $R$-charges $(\Delta_1 ,\cdots ,\Delta_{N_f})$.
Applying the localization method \cite{Pestun:2007rz} ,
the $S^3$ partition function of this theory is given by \cite{Kapustin:2009kz}
\begin{\eq}
Z_{S^3}(g)
=  \int_{-\infty}^\infty d^{|G|} \sigma\
Z_{\rm cl}(\sigma ) Z_{\rm 1loop}(\sigma )  ,
\label{eq:ZS3}
\end{\eq}
where \footnote{
Note that
$Z_{S^3}$ is independent of 
Yang-Mills couplings because of ``$Q$-exactness". 
One can also include FI-term and real mass.
The FI-term gives
a linear function of $\sigma$ to the exponent of $Z_{\rm cl}$ 
and does not change any results in this paper qualitatively.
An effect of the real mass is a constant shift in $Z_{\rm 1loop}$.
While this does not spoil our main conclusion \eqref{eq:conclusion},
the real mass shifts locations of poles in Borel plane and
affects Borel summability along $\mathbb{R}_+$.
}
\begin{\eqa}
&& Z_{\rm cl}(\sigma )
= \exp{\Bigl[  \sum_{p=1}^n \frac{i\sgn (k_p )}{g_p} {\rm tr} (\sigma^{(p)} )^2  \Bigr]} ,\NN\\
&& Z_{\rm 1loop}(\sigma ) 
= \frac{\prod_{\alpha \in {\rm root}_+ }
 4\sinh^2{(\pi \alpha \cdot \sigma )}  }
   {\prod_{m=1}^{N_f} \prod_{\rho_m \in \mathbf{R_m}}  
   s_1 \left( \rho_m \cdot \sigma -i(1-\Delta_m ) \right)} ,\NN\\ 
&& s_b (z)
= \prod_{m=0}^\infty \prod_{n=0}^\infty
\frac{mb +nb^{-1}+Q/2 -iz}{mb +nb^{-1}+Q/2 +iz} .
\end{\eqa}
The parameter $g_p$ is proportional to $1/|k|$.
Now we are interested in small-$g_p$ expansion of $Z_{S^3}(g)$:
\begin{\eq}
Z_{S^3} (g)
= \sum_{\{ \ell_p \} =0}^\infty c_{\ell_1 ,\cdots ,\ell_n} \prod_{p=1}^n g_p^{\frac{{\rm dim}(G_p )}{2} +\ell_p} .
\end{\eq}
We will see that
the perturbative series are usually non-Borel summable along $\mathbb{R}_+$
but always Borel summable 
along negative (positive) imaginary axis for $k_p >0$ ($k_p <0$).

\subsubsection*{$U(N)_k$ adjoint SQCD}
For simplicity of explanations,
we begin with the 3d $\mathcal{N}=2$ $U(N)_k$ SQCD 
with $N_f$ fundamental (R-charge $\Delta_f$),
$\bar{N}_f$ anti-fundamental (R-charge $\bar{\Delta}_f$) and
$N_a$ adjoint chiral multiplets (R-charge $\Delta_a$).
We will discuss general case later.
The $S^3$ partition function of this theory is 
\begin{\eqa}
Z_{\rm SQCD}
&=&  \int_{-\infty}^\infty d^N \sigma\ 
  \prod_{j=1}^N
 e^{\frac{i\sgn (k)}{g} \sigma_j^2}
   \frac{s_1^{\bar{N}_f} \left( \sigma_j +i(1-\bar{\Delta}_f )  \right)}
        {s_1^{N_f} \left(  \sigma_j -i(1 -\Delta_f ) \right)}     \NN\\
&& \times  
  \frac{\prod_{i<j}  4\sinh^2{(\pi (\sigma_i -\sigma_j ) )} }
      {\prod_{i,j} s_1^{N_a} \left( \sigma_i -\sigma_j -i(1-\Delta_a ) \right)} .
\label{eq:SQCD}
\end{\eqa} 
Now we apply the technique in \cite{Honda:2016mvg} to this
and
investigate properties of the small-$g$ expansion of $Z_{\rm SQCD}$.
To do this,
let us make the following change of variables 
\begin{\eq}
\sigma_i = \sqrt{\tau} \hat{x}_i ,
\end{\eq}
where $\hat{\mathbf{x}} =(\hat{x}_1 ,\cdots ,\hat{x}_N )$ is the unit vector
spanning unit $S^{N-1}$.
Then we rewrite the partition function as
\begin{\eqa}
 Z_{\rm SQCD}
&=&  \int_0^\infty d\tau \ e^{\frac{i\sgn (k)}{g}\tau} f (\tau ) \NN\\
&=& i\sgn (k) \int_0^{-i\sgn (k)\infty} dt\ e^{-\frac{t}{g}} f (i \sgn (k)t) ,
\label{eq:FourierSQCD}
\end{\eqa} 
where
\begin{\eqa}
&& f (\tau ) 
= \frac{\tau^{\frac{N^2}{2} -1}}{2} \int_{S^{N-1}} d^{N-1}\hat{x}\ 
 h (\tau ,\hat{x}) ,\NN\\
&&    h (\tau ,\hat{x}) 
= \frac{Z_{\rm VdM}(\hat{x} )  Z_{\rm 1loop}(\sqrt{\tau}\hat{x} ) }
    {Z_{\rm VdM}(\sqrt{\tau}\hat{x}  ) } ,\NN\\
&& Z_{\rm VdM}(\sigma )
= \prod_{\alpha \in {\rm root}_+ } (\pi \alpha\cdot \sigma )^2  .
\label{eq:ft}
\end{\eqa}
Note that
\eqref{eq:FourierSQCD} is similar to the Borel resummation formula \eqref{eq:Borel}
with the direction $\theta =-\pi \sgn (k)/2$.
Therefore
one might wonder 
whether $f(\tau )$ is related to the Borel transformation of 
the original perturbative series.
This question is technically equivalent to
whether $f(\tau )$ consists purely of convergent power series of $\tau$
and it is very nontrivial in general. 

Nevertheless
we can indeed prove in a similar way to \cite{Honda:2016mvg} that
$f(\tau )$ has the following relation to the Borel transformation
\begin{\eq}
i\sgn (k) f(\tau ) = \mathcal{B}Z_{\rm SQCD} (-i\sgn (k) \tau) ,
\label{eq:statement}
\end{\eq}
where $\mathcal{B}Z_{\rm SQCD} (t)$ is the Borel transformation of 
the small-$g$ expansion of $Z_{\rm SQCD}$.
Here we just write down an outline of the proof 
(see appendix for details):
(I) We show uniform convergence of the small-$\tau$ expansion of $h(\tau ,\hat{x})$.
(II) The uniform convergence tells us that
$h(\tau ,\hat{x})$ is 
the same as analytic continuation of the convergent series and
we can exchange the order of the power series expansion of $h(\tau ,\hat{x})$
and the integration over $\hat{x}$.
(III) 
The integral transformation \eqref{eq:FourierSQCD}
guarantees that 
the coefficient of the perturbative series of $f(\tau )$ 
at $\mathcal{O}(\tau^{\frac{N^2}{2}+\ell -1} )$ 
is given by $(-i\sgn (k))^\ell c_\ell /\Gamma (\frac{N^2  +\ell}{2})$ \footnote{
Strictly speaking,
we consider first ${\rm Im}(g)=\epsilon \sgn (k)$ 
and take $\epsilon\rightarrow +0$.
This prescription is 
usually adopted in
perturbative computation of CS-type matrix models
by using the Gaussian matrix model. 
}.
Thus we conclude 
\begin{\eq}
Z_{\rm SQCD}
= \int_0^{-i\sgn (k)\infty} dt\ e^{-\frac{t}{g}}
   \mathcal{B}Z_{\rm SQCD} (t) .
\label{eq:conclusionSQCD}
\end{\eq}
Since the Borel transformation does not have
singularities along the integral contour \footnote{
In our convention,
branch cut of $\sqrt{z}$ is along $\mathbb{R}_-$.
},
the small-$g$ expansion of $Z_{\rm SQCD}$ is Borel summable
along the direction $\theta =-\sgn (k)\pi/2$.
Eq.~\eqref{eq:conclusionSQCD} also tells us that
the Borel resummation with this direction
gives the exact result.

\begin{figure}[t]
\begin{center}
\includegraphics[width=6.5cm]{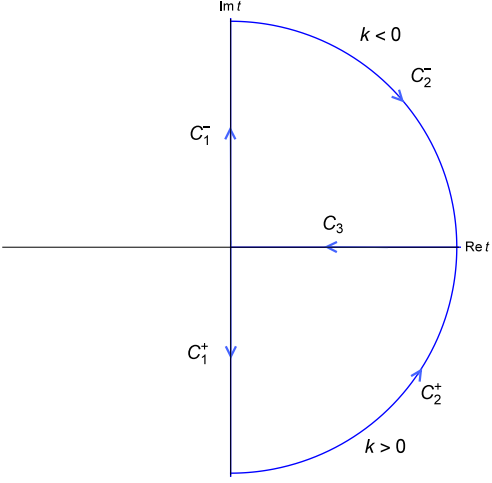}
\end{center}
\caption{
The integral contour,
which relates the Borel resummation along the imaginary axis
to the one along the real positive axis.
}
\label{fig:contour}
\end{figure}
When does the pertubative series 
become Borel summable 
along $\mathbb{R}_+$?
Since $t\in\mathbb{R}_+$ corresponds
to $\sigma_j \in (-e^{\frac{\pi i}{4}\sgn (k) } \infty ,e^{\frac{\pi i}{4}\sgn (k) } \infty )$ in \eqref{eq:SQCD},
a sufficient condition for this is
absence of singularities in the one-loop determinant along this line, namely $N_a =0$ (or $\Delta_a =1$).
Next we ask
when the perturbative series is Borel summable along $\mathbb{R}_+$,
how this is related to the exact result.
To answer this question,
we need to change the integral contour to $\mathbb{R}_+$
as in fig. \ref{fig:contour}.
There is a subtlety on this,
which is related to CS level shift 
coming from integration over massive fermions (see e.g.~\cite{Kao:1995gf}).
When the integral variables $\sigma$ 
are very large,
the contribution from chiral multiplet becomes
\begin{\eq}
\frac{1}{s_1 (\sigma -i(1-\Delta ) )}  
= \exp{\Biggl[  \frac{i\pi\sgn (\sigma )}{2} \sigma^2 +\mathcal{O}(|\sigma |) \Biggr]  } .  
\end{\eq}
This effectively shifts the CS level by $\sgn (\sigma )/2$ and
the shift in the adjoint SQCD is 
totally $\sgn (\sigma_j )(N_f -\bar{N}_f)/2$.
Hence the contribution from $C_2^\pm$ disappears 
for $|k|>|N_f -\bar{N}_f |/2$.
If we consider this region, then we find
\begin{\eq}
Z_{\rm SQCD}
=\left( \int_0^{\infty }dt +\oint_C dt \right)
e^{-\frac{t}{g}} \mathcal{B}Z_{\rm SQCD} (t) ,
\end{\eq}
where the integral contour $C$ is 
$C=C_1^+ +C_2^+ +C_3$ for $k>0$
and $C=C_1^- +C_2^- +C_3$ for $k<0$.
Thus the Borel resummation along $\mathbb{R}_+$ 
gives the exact result when the second term is zero.
A sufficient condition for this is again $N_a =0$. 

It is worth looking at $N_f =\bar{N}_f =N_a =0$ case,
which corresponds to the $\mathcal{N}=2$ CS theory without chiral multiplets
and the same as the pure CS theory up to level shift.
Since $Z_{\rm 1loop}$ does not have poles for this case,
the Borel transformation also does not have any poles.
This reflects the fact that 
the perturbative series in the pure CS theory
is convergent.

\subsubsection*{General 3d $\mathcal{N}=2$ CS matter theory}
Extension to general 3d $\mathcal{N}=2$ CS matter theory is straightforward.
First we  insert delta function constraint $\Delta (\sigma )$ to the integrand \footnote{
If 
$G_p$ is $SU(N)$, we insert $\delta (\sum_{j=1}^N \sigma^{(p)}_j )$.
}
such that the following coordinate spans sphere with radius $\sqrt{\tau_p}$ 
\begin{\eq}
\sigma_i^{(p)} = \sqrt{\tau_p} \hat{x}_i^{(p)} .
\label{eq:coord}
\end{\eq}
Then the partition function again takes 
the form of \eqref{eq:FourierSQCD}
extended to multi-variables:
\begin{\eqa}
 Z_{S^3}
&=& \int_0^\infty d^n \tau\ 
 e^{\sum_{p=1}^n \frac{i\sgn (k_p ) }{g_p}\tau_p} f(\tau ) \NN\\
&=&\Biggl[  \prod_{p=1}^n i\sgn (k_p) \int_0^{-i\sgn (k_p) \infty} dt_p  
 e^{-\frac{t_p }{g_p}} \Biggr] f(i\sgn (k)t ) , \NN\\
\label{eq:Laplace}
\end{\eqa} 
where
\begin{\eqa}
&& f (\tau ) 
=\frac{\tau^{\frac{{\rm dim}(G)}{2}-1}}{2^n}  
\int_{\rm sphere}d\hat{x}\ 
\Delta (\hat{x}) h (\tau ,\hat{x}) ,\NN\\
&& h (\tau ,\hat{x}) 
=\left.  \frac{Z_{\rm VdM}(\hat{x} )  Z_{\rm 1loop}(\sigma ) }{Z_{\rm VdM}(\sigma ) }
\right|_{\sigma_i^{(p)} = \sqrt{\tau_p} \hat{x}_i^{(p)}} ,\NN\\
&& \tau^{\frac{{\rm dim}(G)}{2}-1} 
=\prod_{p=1}^n \tau_p^{\frac{{\rm dim}(G_p )}{2}-1} .
\end{\eqa}
We can always prove that
$f(\tau )$ is 
related to the Borel transformation of the original perturbative series as
\begin{\eq}
 \Biggl[ \prod_{p=1}^n i\sgn (k_p ) \Biggr] f( \{ \tau_p \} ) 
= \mathcal{B}Z_{S^3} \left( \{ -i\sgn (k_p ) \tau_p \} \right) ,
\end{\eq}
since small-$\tau_p$ expansion of $h(\tau ,\hat{x})$ is uniform convergent
if $Z_{S^3}$ is well-defined.
This immediately leads us to 
\begin{\eq}
Z_{S^3} (g)
=\Biggl[ \prod_{p=1}^n \int_0^{-i\sgn (k_p )\infty} dt_p  e^{- \frac{t_p}{g_p}} \Biggr]
   \mathcal{B}Z_{S^3} (t) ,
\label{eq:main}
\end{\eq}
which is generalization of \eqref{eq:conclusionSQCD}.

A sufficient condition for Borel summability along $\mathbb{R}_+$
is again absence of singularities along $\sigma_j \in (-e^{\frac{\pi i}{4}\sgn (k) } \infty ,e^{\frac{\pi i}{4}\sgn (k) } \infty )$
in $Z_{\rm 1loop}$.
When the perturbative series is Borel summable along $\mathbb{R}_+$
and ``level shift'' is not so very large,
we obtain
\begin{\eq}
Z_{S^3}(g)
=\Biggl[ \prod_{p=1}^n \Bigl( \int_0^{\infty }dt_p +\oint_C dt_p \Bigr)
e^{-\frac{t_p}{g_p}} \Biggr] \mathcal{B}Z_{S^3} (t) .
\label{eq:relation_usual}
\end{\eq}
If the second term is zero,
the Borel resummation along $\mathbb{R}_+$ is the same as the exact result.
In the rest of this paper,
we prove our main result for various observables: 
\begin{\eq}
\mathcal{O} (g)
=\Biggl[ \prod_{p=1}^n \int_0^{-i\sgn (k_p )\infty} dt_p  e^{- \frac{t_p}{g_p}} \Biggr]
   \mathcal{B}\mathcal{O} (t) .
\label{eq:mainO}
\end{\eq}

\subsection{Other observables}
\subsubsection*{Supersymmetric Wilson loop}
We can 
generalize the above considerations to other observables.
Let us begin with the Wilson loop
\begin{\eq}
W_{\mathbf{R}}(C)
={\rm tr}_{\mathbf{R}}
P\exp{\Biggl[ \oint_C ds (iA_\mu \dot{x}^\mu +\sigma |\dot{x}| ) \Biggr]} ,
\label{eq:Wilson}
\end{\eq}
with the adjoint scalar $\sigma$ in vector multiplet
The Wilson loop preserves two supercharges 
when the contour $C$ is the great circle of $S^3$. 
Applying the localization method,
VEV of the Wilson loop is given by 
\begin{\eq}
\langle W_{\mathbf{R}}({\rm Circle})  \rangle
=\langle {\rm tr}_{\mathbf{R}} e^\sigma \rangle_{\rm M.M.} ,
\label{eq:Wilson}
\end{\eq}
where $\langle \cdots \rangle_{\rm M.M.}$ denotes VEV in the matrix model \eqref{eq:ZS3}.
This is just linear combination of
exponential function of $\sigma$ and
we can obviously write the Wilson loop as in
\eqref{eq:mainO}.

\subsubsection*{Bremsstrahrung function in SCFT on $\mathbb{R}^3$}
Bremsstrahrung function $B$
determines
an energy radiated by accelerating quarks in small velocities
as $E=2\pi B\int dt \dot{v}^2$.
It was conjectured that
$B$
in 3d $\mathcal{N}=2$ superconformal theory
is given by \cite{Lewkowycz:2013laa}
\begin{\eq}
B(g)
=\frac{1}{4\pi^2}\left. \frac{\del}{\del b}
\log \langle {\rm tr} e^{ba} \rangle_{\rm M.M.} \right|_{b=1} ,
\end{\eq}
which is technically derivative of the Wilson loop in fundamental representation
with winding number $b$.
As in the Wilson loop,
we can also rewrite $B (g)$ as in \eqref{eq:mainO}.

\subsubsection*{Two-point function of $U(1)$ flavor symmetry currents in SCFT on $\mathbb{R}^3$}
Next we consider
two-point function 
of the $U(1)$ flavor symmetry current $j_\mu^a$
for superconformal case.
The 3d conformal symmetry fixes the two-point function as
\begin{\eq}
\langle j_a^\mu (x) j_b^\nu (0) \rangle
=\frac{\tau_{ab}}{16\pi^2}
 (\delta^{\mu\nu}\del^2 -\del^\mu \del^\nu ) \frac{1}{x^2} 
+\frac{i\kappa_{ab}}{2\pi}
 \epsilon^{\mu\nu\rho} \del_\rho \delta^{(3)}(x) ,
\end{\eq}
where $\tau_{ab}(g)$ and $\kappa_{ab}(g)$
are independent of $x$ but nontrivially dependent on parameters.
We can exactly compute $\tau_{ab}(g)$ and $\kappa_{ab}(g)$
by the localization \cite{Closset:2012vg}.
This is generated by the $S^3$ partition function $Z_{S^3}(m,g)$
deformed by real mass $\{m_a \}$ associated with the $U(1)$ symmetries:
\begin{\eqa}
&& \tau_{ab} (g)
= -\frac{2}{\pi^2}{\rm Re}\left[ 
\frac{1}{Z_{S^3}(0,g)} \frac{\del^2 Z_{S^3}(m,g) }{\del m_a \del m_b}  
  \right]_{ \{m_a \} =0} ,\NN\\
&& \kappa_{ab}(g) = \frac{1}{2\pi} {\rm Im}\left[ 
\frac{1}{Z_{S^3}(0,g)} \frac{\del^2 Z_{S^3}(m,g) }{\del m_a \del m_b}  
 \right]_{ \{m_a \} =0} .
\end{\eqa}
Repeating the argument on $Z_{S^3}$,
we can show that
$\tau_{ab}(g)$ and $\kappa_{ab}(g)$ satisfy \eqref{eq:mainO}.

\subsubsection*{Partition function and Wilson loop on Squashed $S^3$}
Let us consider partition function on squashed sphere $S^3_b$
with the squashing parameter $b$ \footnote{
Although there are many choices of $S_b^3$,
we have the same partition function \cite{Hama:2011ea} 
as long as it is one-parameter deformation of the round $S^3$ 
keeping SUSY \cite{Closset:2013vra} (see also \cite{Imbimbo:2014pla}).
}.
This has a simple relation 
to supersymmetric Renyi entropy \cite{Nishioka:2013haa}.
Only difference from $Z_{S^3}$ in localization formula 
is the one-loop determinant \cite{Hama:2011ea}: 
\begin{\eq}
 Z_{\rm 1loop}(\sigma ) 
= \frac{\prod_{\alpha \in {\rm root}_+ }
 4\sinh{(\pi b\alpha \cdot \sigma )} 
 \sinh{(\pi b^{-1}\alpha \cdot \sigma )} }
   {\prod_{m=1}^{N_f} \prod_{\rho_m \in \mathbf{R_m}}  
   s_b \left(  \rho_m \cdot \sigma -\frac{iQ}{2}(1-\Delta_m ) \right)} , 
\end{\eq} 
with $Q=b+b^{-1}$.
Note that
the partition function is ill-defined
when one of $m_1 b+m_2 b^{-1}$ ($m_{1,2}\in\mathbb{Z}$) is purely imaginary. 
Otherwise
we arrive at the same conclusion \eqref{eq:mainO}
by a similar argument.
An important difference from the round sphere case is that
the poles of the Borel transformation rotate 
as varying the argument of $b$ and
hit the integral contour of \eqref{eq:mainO}
when the partition function becomes ill-defined.

One can also consider SUSY Wilson loop on ellipsoid
constructed in \cite{Tanaka:2012nr}.
This Wilson loop has a topology of torus knot 
when $b^2$ is rational number.
As in \eqref{eq:Wilson},
localization formula of the Wilson loop
is VEV of ${\rm tr}_{\mathbf{R}} e^\sigma$
in the matrix model of the squashed sphere.
Hence the Wilson loop can be also written as in \eqref{eq:mainO}.

\subsubsection*{Two point function of stress tensor in SCFT on $\mathbb{R}^3$}
In 3d CFT,
two point function of canonically normalized stress tensor at separate points
takes the form \cite{Osborn:1993cr}
\begin{\eq}
\langle T_{\mu\nu}(x) T_{\rho\sigma}(0) \rangle
=\frac{c_T}{64} (P_{\mu\rho}P_{\nu\sigma} +P_{\nu\rho}P_{\mu\sigma} -P_{\mu\nu}P_{\rho\sigma}) \frac{1}{16\pi^2 x^2},
\end{\eq}
where $P_{\mu\nu}=\delta_{\mu\nu}\del^2 -\del_\mu \del_\nu$
\footnote{In this normalization
$c_T =1$ for one free real scalar and Majorana fermion. 
}.
The coefficient $c_T (g)$
can be computed by $Z_{S_b^3}$ as \cite{Closset:2012ru}  
\begin{\eq}
c_T (g)
= -\frac{32}{\pi^2}{\rm Re} \left[ 
 \frac{1}{Z_{S^3}(g)} \frac{\del^2 Z_{S_b^3}(g)}{\del b^2}
 \right]_{b=1} .
\end{\eq}
By a similar argument,
\eqref{eq:mainO} holds also for $c_T (g)$.

\subsubsection*{Partition function on squashed lens space}
Suppose orbifold of bi-axially squashed sphere: $S_b^3 /\mathbb{Z}_n$
\footnote{
Regarding $S^3$ as $S^1$-bundle over $S^2$,
this is roughly a rescale of the $S^1$-fibre.
}.
Gauge theory on the lens space has degenerate vacua
specified by
$m = \frac{n}{2\pi} \oint A$ ,
where the integral contour is an element of $\pi_1 (S_b^3 /\mathbb{Z}_n )$.
Therefore partition function on this space
is decomposed as
\begin{\eq}
Z_{S_b^3 /\mathbb{Z}_n}
=\sum_m Z_{S_b^3 /\mathbb{Z}_n}^{(m)} .
\end{\eq}
The localization method tells us that
$Z_{S_b^3 /\mathbb{Z}_n}^{(m)}$ is expressed as in \eqref{eq:ZS3}
with the different one-loop determinant \cite{Imamura:2012rq}
\begin{\eq}
 Z_{\rm 1loop}^{(m)} 
= \frac{\prod_{\alpha \in {\rm root} }
 s_{b,\alpha (m)} (\alpha \cdot \sigma -iQ/2 ) } 
   {\prod_{f=1}^{N_f} \prod_{\rho_f \in \mathbf{R_f}}  
   s_{b,\rho_f (m)} \left( \rho_f \cdot \sigma -iQ(1-\Delta_f )/2  \right)} ,
\end{\eq} 
where
\begin{\eqa}
s_{b,h}(z)
&=& \prod_{p=0}^{n-1} 
s_b \left( \frac{z}{n} +ib\langle p\rangle_n
   +ib^{-1}\langle p+h \rangle_n  \right) ,\NN\\
\langle m \rangle_n
&=& \frac{1}{n} \left( [m]_n +\frac{1}{2} \right) -\frac{1}{2} .
\end{\eqa}
One can prove \eqref{eq:mainO} for $Z_{S_b^3 /\mathbb{Z}_n}^{(m)}$
by the same argument as the squashed $S^3$ partition function.

\section{Discussions}
We have studied the perturbative series 
in general 3d $\mathcal{N}=2$ SUSY CS matter theory.
We have proven that
the perturbative series are 
Borel summable along negative (positive) imaginary axis 
for positive (negative) CS levels
and
the Borel resummations along this direction are the same as the exact results
for various observables.
Thus we conclude that
the Borel resummations of this direction are correct ways of resumming the perturbative series.
We have found that
this structure is already hidden in the localization formula.

We have found that
the perturbative series
are usually not Borel summable along $\mathbb{R}_+$
due to the singularities in the Borel transformations.
It is interesting to find
physical interpretations of the singularities.
Technically the singularities come
from poles in one-loop determinant of chiral multiplets.
It is known 
in the context of factorization \cite{Pasquetti:2011fj} that
the poles for the squashed $S^3$ partition function
correspond to Higgs branch solutions.
Hence we expect that
the singularities are related to 
such semiclassical solutions.
It would be nice if one can make it clearer.

While the sufficient condition for Borel summability along $\mathbb{R}_+$
is absence of singularities along $\sigma_j \in (-e^{\frac{\pi i}{4}\sgn (k) } \infty ,e^{\frac{\pi i}{4}\sgn (k) } \infty )$
in $Z_{\rm 1loop}$,
there should be many theories,
which do not satisfy this condition but are Borel summable 
along $\mathbb{R}_+$. 
One of such examples
is the $S^3$ partition function of 
3d $\mathcal{N}=6$ superconformal theory 
(ABJM theory \cite{Aharony:2008ug})
with $U(2)\times U(2)$ gauge group \cite{Russo:2012kj}.
It is very important
to find necessary or more sufficient conditions
for Borel summability along $\mathbb{R}_+$.
Since we have shown 
Borel summability along $\mathbb{R}_+$
for 4d and 5d theories with eight supercharges in \cite{Honda:2016mvg},
it might be natural to expect that
pertuabative series in 3d $\mathcal{N}=4$ CS matter theories
are Borel summable along $\mathbb{R}_+$.

For theories describing M2-branes,
the CS levels are not completely independent of each other and
satisfy $\sum_{p=1}^n k_p =0$.
While our analysis includes such M2-brane theories as special cases,
we could directly discuss these cases.
One of subtleties here is that
if we take $\sum_{p=1}^n k_p =0$ at first in our argument,
then integral domain of $\hat{x}$ in \eqref{eq:coord} becomes non-compact.
It is very nice
if one can overcome the subtleties.

In the planar limit,
we expect that
the perturbative series become convergent \cite{'tHooft:1982tz}
and hence Borel summable along positive real axis.
To be consistent with this,
the second term in \eqref{eq:relation_usual}
should be suppressed in $1/N$-expansion.
It is illuminating if one can explicitly prove this statement.
This would be also related to
a simple connection between the planar limit and
``M-theory limit" discussed in \cite{Azeyanagi:2012xj}.

Recently it was discussed that
some SUSY CS matter theories
exhibit phase transitions 
as varying real masses or FI-parameters \cite{Barranco:2014tla}.
Since real masses shift poles of $Z_{\rm 1loop}$,
these also shift poles in Borel plane.
In general this effect may change directions of Borel summability and
be related to the phase transitions.

Finally,
although we know localization formula for vortex loop \cite{Drukker:2012sr},
we have not discussed perturbative series of the vortex loop.
Technically the localization formula for the vortex loop
is like the $S^3$ partition function with a different integral contour
and 
probably we need to think of it more carefully.

\vspace{.5cm}
\acknowledgments
We thank Zohar Komargodski and Jorge G.~Russo 
for helpful comments on the draft.

\appendix
\section{Proof of (10)}
\label{app:detail}
Here
we explicitly prove
\eqref{eq:statement} 
as in \cite{Honda:2016mvg}.
For this purpose,
first we prove
uniform convergence of the small-$\tau$ expansion of $h(\tau ,\hat{x})$.
Let us rewrite $h(\tau ,\hat{x})$
in a convenient form for the small-$\tau$ expansion.
By using
\[
 \frac{\sinh{\pi x}}{\pi x}
=  \prod_{n=1}^\infty \left( 1 +\frac{x^2}{n^2} \right) ,\quad
 s_1 (z)
=\prod_{n=1}^\infty \left( \frac{n-iz}{n+iz} \right)^n ,
\]
we find that
the small-$\tau$ expansion is generated by
\begin{\eqa}
&& 2^{N^2 -N}Z_{\rm VdM} (\hat{x}) 
\exp \Biggl[ -2\sum_{i<j} \sum_{\ell =1}^\infty \frac{(-\tau )^\ell \zeta (2\ell )}{\ell} (\hat{x}_i -\hat{x}_j )^{2\ell} \NN\\
&& -N_a \sum_{i,j} \ln\tilde{s}_1 
  (\sqrt{\tau}(\hat{x}_i -\hat{x}_j ) -i(1-\Delta_a )  )  \NN\\
&& -N_f \sum_{j} \ln\tilde{s}_1 
   ( \sqrt{\tau}\hat{x}_j -i(1-\Delta_f ) )  \NN\\
&& -\bar{N}_f \sum_{j} \ln\tilde{s}_1
    ( -\sqrt{\tau}\hat{x}_j -i(1-\bar{\Delta}_f ) )
 \Biggr] ,
\label{eq:generate}
\end{\eqa}
where $\tilde{s}_1 (x)$
is a generating function of small-$x$ expansion of $s_1 (x)$:
\begin{\eq}
\tilde{s}_1 (x)
= \exp{\Biggl[ -2ix \sum_{\ell =0}^\infty \frac{\zeta (2\ell )}{2\ell +1} (-x^2 )^\ell  \Biggr]} .
\end{\eq}
To show uniform convergence of the small-$\tau$ expansion,
we apply Weierstrass's M-test,
which ask if one can find
a sequence $\{ M_\ell \}$ satisfying
$| h_\ell (\hat{x})| < M_\ell $ and 
$\sum_{\ell =0}^\infty M_\ell  <\infty $
for fixed $\tau$.
Indeed we can easily construct such a series.
For instance, 
since $\zeta (\ell\geq 2)<2$ and $\hat{x}\leq 1$,
a generating function $\bar{h}(\tau )$ of $M_\ell$
can be obtained by the replacement in \eqref{eq:generate}:
\begin{\eqa *}
(-1)^{\ell +1}\zeta (2\ell )(\hat{x}_i -\hat{x}_j )^{2\ell}
&\rightarrow & 2 ,\NN\\
-\ln\tilde{s}_1 (\sqrt{\tau}\hat{x} -i (1-\Delta )  ) 
&\rightarrow & 4 \sum_{\ell =0}^\infty
 \frac{(\sqrt{\tau} +|1-\Delta |)^{2\ell +1}}{2\ell +1}      \NN\\
&=& 2\log{ \frac{1+\sqrt{\tau} +|1-\Delta |}{1-\sqrt{\tau} -|1-\Delta | } }
, \NN\\
\end{\eqa *}
which leads us to
\begin{\eqa *}
&&\bar{h}(\tau ) 
= \frac{ 2^{N^2 -N}Z_{\rm VdM} (\hat{x}) }{(1-\tau )^{2(N^2 -N)}} 
\left( \frac{1 +\sqrt{\tau} +|1-\Delta_a |}{1-\sqrt{\tau} -|1-\Delta_a | } \right)^{2N^2 N_a} \NN\\
&& \left( \frac{1+\sqrt{\tau} +|1-\Delta_f |}{1-\sqrt{\tau} -|1-\Delta_f | } \right)^{2N N_f}
\left( \frac{1+\sqrt{\tau} +|1-\bar{\Delta}_f |}
            {1-\sqrt{\tau} -|1-\bar{\Delta}_f | } \right)^{2N \bar{N}_f} .
\end{\eqa *}
Thus
the small-$\tau$ expansion of $h(\tau ,\hat{x})$
is uniform convergent.
This implies that
$h(\tau ,\hat{x})$ is 
analytic continuation of the convergent series, and
we can exchange the power series expansion of $h(\tau ,\hat{x})$
and the integration over $\hat{x}$.
Therefore $f(\tau )$ is also identical to
an analytic continuation of the convergent series.
Finally the integral transformation 
\eqref{eq:FourierSQCD}
gives \eqref{eq:statement}.

 
\end{document}